\documentclass[letter,twocolumn]{jpsj3}
\usepackage{txfonts}
\usepackage{graphicx}
\usepackage{color}
\usepackage{amsmath}
\usepackage{amssymb}
\usepackage{array}
\usepackage{wasysym}
\usepackage{braket}
\usepackage{nidanfloat}
\usepackage{ulem} 

\title{Inelastic Neutron Scattering Study of the Spin Dynamics\\ in the Breathing Pyrochlore System LiGa$_{0.95}$In$_{0.05}$Cr$_4$O$_8$}

\author{Yu Tanaka$^1$\thanks{footnote}, Rafal Wawrzy\'{n}czak$^2$, Manh Duc Le$^3$, Tatiana Guidi$^3$, Yoshihiko Okamoto$^4$, Takeshi Yajima$^1$, Zenji Hiroi$^1$, Masashi Takigawa$^1$, and G{\o}ran J. Nilsen$^3$}

\inst{$^1$Institute for Solid State Physics, University of Tokyo, Kashiwa, Chiba 277-8581, Japan \\ $^2$Institut Laue-Langevin, CS 20156, Cedex 9, 38042 Grenoble, France \\ $^3$ISIS Facility, Rutherford Appleton Laboratory-STFC, Chilton, Didcot OX11 0QX, United Kingdom \\ $^4$Department of Applied Physics, Nagoya University, Nagoya 464-8603, Japan}

\abst{The $A$-site ordered chromate spinels LiGa$_{1-x}$In$_{x}$Cr$_4$O$_8$ host a network of size-alternating spin-3/2 Cr$^{3+}$ tetrahedra known as a ``breathing'' pyrochlore lattice. For the $x=0.05$ composition, the complex magneto-structural ordering observed in the parent $x=0$ material is replaced by a single transition at $T_f=11$~K, ascribed to the collinear nematic order caused by strong spin-lattice coupling. We present here an inelastic neutron scattering study of the spin dynamics in this composition. Above $T_f$, the dynamical scattering function $S(Q,E)$ is ungapped and quasi-elastic, similar to undoped LiGaCr$_4$O$_8$. 
Below $T_f$, the spectral weight splits between a broad inelastic feature at $5.8$~meV and toward the elastic line. The former feature can be ascribed to spin precessions within antiferromagnetic loops, lifted to finite energy by the effective biquadratic spin-lattice term in the spin Hamiltonian.} 

\begin{document}
\maketitle

When magnetic frustration is combined with strong spin-lattice coupling, a range of possible magneto-structural behaviors result \cite{moessner2000}, including nematic transitions \cite{wawrzynczak2017}, magnetization plateaus, and localized spin excitations \cite{lee2000,tomiyasu2011}. An ideal arena for exploring this interplay is provided by the chromate spinels, $A^{2+}$Cr$_2^{3+}$O$_4$. Here, the frustration originates from the corner-sharing pyrochlore network of Cr$^{3+}$ ($S=3/2$) tetrahedra on the $B$-site, while the strong spin-lattice coupling arises from the direct overlap of Cr$^{3+}$ $d$-orbitals on adjacent sites. A common starting point to understand the low-temperature physics of the chromate spinels is the so-called bilinear-biquadratic model \cite{tchernyshyov02,shannon10},
 \begin{equation}
	\mathcal{H}=J\sum_{i,j}\mathbf{S}_{i}\cdot\mathbf{S}_{j}+b\sum_{i,j}(\mathbf{S}_{i}		\cdot\mathbf{S}_{j})^{2}
	\label{bbm}
\end{equation}   
    where the classical nearest-neighbor Heisenberg Hamiltonian is extended with an effective biquadratic term, $b(\mathbf{S}_i\cdot \mathbf{S}_j)^2$ (where $b$ is a coupling constant, and $\mathbf{S}_{i,j}$ are classical spins). This term, generated by spin-lattice coupling to local distortions, lifts some of the degeneracy of the ground state manifold by selecting collinear or coplanar spin configurations. Although the bilinear-biquadratic model ignores the long-range interactions which eventually cause magneto-structural order in many chromate spinels, it is able to successfully describe both the short-range spin correlations and the magnetic phase diagram of virtually every member of the family. 
    It has not, however, yet been applied to the unusual magnetic excitation spectra in the ordered phases of MgCr$_2$O$_4$ \cite{tomiyasu2013}, HgCr$_2$O$_4$ \cite{tomiyasu2011}, and ZnCr$_2$O$_4$ \cite{ratcliff02}; these are characterized by weak spin-wave branches and sharp, non-dispersive inelastic bands, with wave vector dependences characteristic of small spin clusters\cite{tomiyasu2011,tomiyasu08}. While bands assigned to both hexamer and heptamer clusters are observed in MgCr$_2$O$_4$ and HgCr$_2$O$_4$, only the former are seen in ZnCr$_2$O$_4$. Due to the complexity of the low-temperature magneto-structural orders in these materials, the connection between the structure and the apparent spin cluster excitations is unclear.
    
    This link is more evident, at least at high temperature, in the so-called ``breathing'' pyrochlore spinels $A^+A^{\prime 3+}$Cr$_4$O$_8$, where the $A$-site is now populated by an ordered arrangement of mono- and trivalent cations. Here, the term ``breathing'' refers to the alternation of Cr$^{3+}$-Cr$^{3+}$ distances and, hence, magnetic exchanges $J$ and $J^\prime$ between adjacent Cr$^{3+}_4$ tetrahedra as a consequence of the order on the $A$-site. The degree of magnetic alternation is quantified by the breathing factor $B_f=J^\prime/J$,  where $B_f\rightarrow 0$ corresponds to isolated tetrahedra and $B_f \rightarrow 1$ to the isotropic pyrochlore lattice. Starting from the former limit, the excitations at $T\sim J$ are localized non-dispersive triplets (and higher multiplets) separated by a spin gap $\Delta$ from the singlet ground state. When $B_f$ is increased, $\Delta$ is suppressed, and the excitations become qualitatively similar to the isotropic case beyond $B_f\sim 0.25$. This simple picture is again complicated by the influence of the spin-lattice coupling and long-ranged terms, which are responsible for the collective excitations in the low-temperature ordered phases. 

LiGaCr$_4$O$_8$ and LiInCr$_4$O$_8$ were rediscovered by Okamoto \textit{et. al.} \cite{okamoto13} as breathing pyrochlore systems, which have $A^+=$ Li$^+$ and $A^{\prime 3+} = $ Ga$^{3+}$/In$^{3+}$. 
The nearest-neighbor magnetic interactions on the small and large tetrahedra are estimated to be $J\sim 50$~K and $J^\prime \sim 30$~K ($B_f\sim 0.6$) for LiGaCr$_4$O$_8$, and $J\sim 60$~K and $J^\prime \sim 6$~K ($B_f\sim 0.1$) for LiInCr$_4$O$_8$ \cite{okamoto13}.
For LiGaCr$_4$O$_8$, the upper magneto-structural transition at $T_u\sim 20$~K results in phase separation into cubic paramagnetic and tetragonal collinear phases. The cubic phase then undergoes another transition at $T_l = 13.8$~K, into a second tetragonal phase, the structure of which has not yet been determined.
As in other chromate spinels, both transitions are first-order. However, the paramagnetic component shows a divergence in the nuclear spin-lattice relaxation rate $1/T_1$ extracted from $^7$Li-NMR, implying proximity to a tricritical point or to a second-order transition to another phase.\cite{tanaka14} 
    
    In this letter, we describe inelastic neutron scattering measurements of the spin excitation spectrum of the ``breathing'' pyrochlore chromate spinel, LiGa$_{0.95}$In$_{0.05}$Cr$_4$O$_8$, where $B_f \sim 0.6$. Previous diffraction measurements indicate that LiGa$_{0.95}$In$_{0.05}$Cr$_4$O$_8$ undergoes a possible second-order transition to a nematic collinear ground state at $T_f=11.1$~K, in accordance with predictions from the bilinear-biquadratic model\cite{wawrzynczak2017}. The excitations at $T>T_f$ are gapless and Lorentzian in form, as is also the case for MgCr$_2$O$_4$, HgCr$_2$O$_4$, and ZnCr$_2$O$_4$. Below $T_f$, the spectral weight shifts to the elastic line and an inelastic feature at ${\sim}5.8$~meV. We identify the latter with spin precession within antiferromagnetic hexagonal spin clusters created by the nematic order, and lifted to finite energy by the biquadratic term. We thus provide, for the first time, a plausible link between the magnetic excitations and magnetic structure. The remaining spectral weight appears to be consistent with collective spin-wave-like excitations.

The powder sample of LiGa$_{0.95}$In$_{0.05}$Cr$_4$O$_8$ was prepared by sintering a stoichiometric mixture of the two end-member compounds LiGaCr$_4$O$_8$ and LiInCr$_4$O$_8$ \cite{okamoto15}. These were in turn prepared by the standard solid-state route \cite{okamoto13}, using starting materials enriched with $^7$Li to reduce neutron absorption. For our inelastic neutron scattering measurements, 8.1 g of powder was packed in an Al sachet, which was rolled into an annulus and loaded into an Al can with $\diameter=45$~mm. The measurements were performed on the MARI direct-geometry time-of-flight chopper spectrometer at the ISIS facility, UK, using incident energies $E_i=10, 16, 25$, and $35$~meV. For all values of $E_i$, the elastic energy resolution was close to $\Delta E/E\sim4.5 \% $. Temperatures between $5$ and $300$~K were accessed using a closed-cycle refrigerator. The diffraction measurements reported in Ref.~\citen{wawrzynczak2017} were performed on the same sample.

    \begin{figure*}[tbh]
    \begin{center}
	\includegraphics[width=0.8\linewidth]{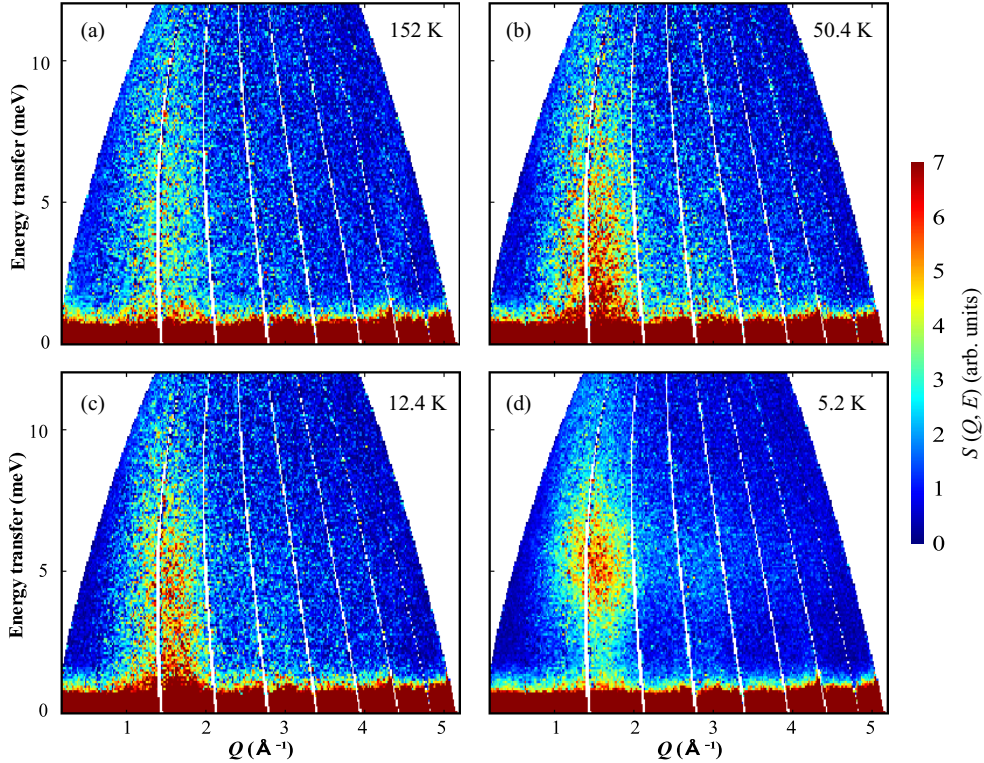}
	\end{center}
	\caption{(Color online) Temperature dependence of $S(Q,E)$ for LiGa$_{0.95}$In$_{0.05}$Cr$_4$O$_8$ recorded with 16~meV incident energy. (a)-(c) are taken at $T>T_f$ and (d) below  $T_f$. Blank patches are due to gaps between detectors.}
		\label{SQW}
	\end{figure*}

\begin{figure}
 		\begin{center}
		\includegraphics[width=7.0cm]{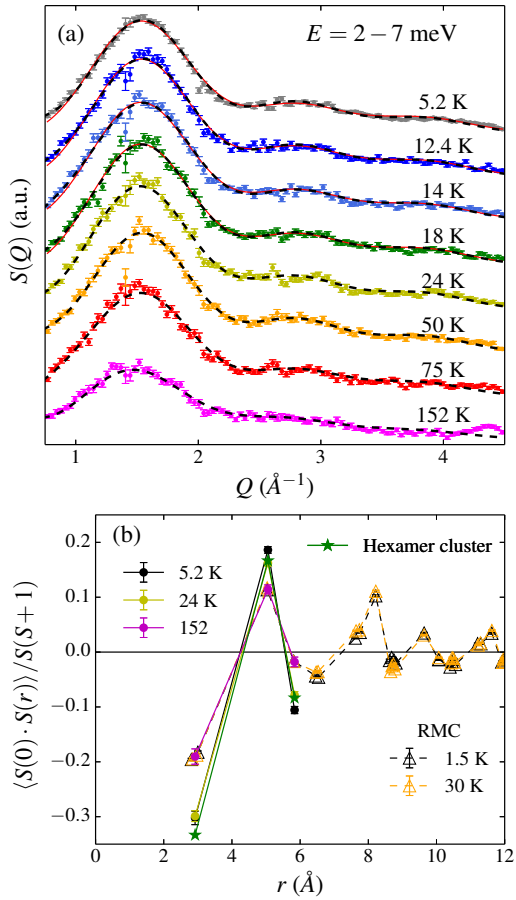}
        \end{center}
		\caption{(Color online) (a) $Q$ dependence of the magnetic scattering $I(Q)$, integrated over the energy range 2-7~meV at different temperatures. Dashed lines are fitting curves calculated from the shell model (Eq.~(\ref{Sq})) with the first three nearest neighbors and flat backgrounds. Solid red lines show the results of structure factor calculations for hexagonal chromium rings at $T<20$~K.
        (b) Real space spin-spin correlation functions $\braket{S(0)\cdot S(r_i)}$ versus $r$. Solid circles are obtained from the fits to the shell model in Fig.~\ref{Qdependence}(a), and the open triangles are obtained by the reverse Monte Carlo (RMC) simulation on the magnetic diffuse scattering observed in the elastic ND measurement. \cite{wawrzynczak2017} Green stars mark the correlations for an isolated hexagonal antiferromagnetic loop (Fig. \ref{hexamer}).}
		\label{Qdependence}
	\end{figure}

The dynamical structure factors $S(Q,E=E_i-E_f)$ measured at four selected temperatures between $5.2$~K and $152$~K are shown in Fig.~\ref{SQW}. We begin with an analysis of the data taken above $T_f$: at $T=152$~K$\gg T_f$, a quasi-elastic rod of scattering extending up to ${\sim}15$~meV is observed. This is characteristic of diffusive spin excitations in the correlated paramagnetic state \cite{conlon2009}, and resembles the $S(Q,E)$ of both LiInCr$_4$O$_8$\cite{nilsen15} and LiGaCr$_4$O$_8$\cite{reportill} at similar temperatures. The intensity of the rod is centered around $1.6$~\AA $^{-1}$, corresponding approximately to the reciprocal space position of the Cr-Cr nearest-neighbor distance. 
Upon cooling to $50.4$~K, intensity builds up near the elastic line, again as in LiGaCr$_4$O$_8$, but in contrast to LiInCr$_4$O$_8$, where the scattering becomes inelastic \cite{nilsen15}. The modulation of the quasi-elastic scattering is also enhanced, indicating the development of longer-ranged spin-spin correlations $\braket{S(0)\cdot S(r)}$, as may be seen in Fig.~\ref{Qdependence}. 

To determine the spatial extent of the correlations, we fit the $Q$-dependence of the scattering integrated between $2$ and $7$~meV ($\simeq S(Q)$ at high temperature) to a shell model [Fig.~\ref{Qdependence}(a)]: 
 \begin{equation}
  S(Q)={f(Q)}^2\sum_{i}\braket{S(0)\cdot S(r_i)}N_i\frac{\sin(Qr_i)}{Qr_i},
  \label{Sq}
\end{equation}
where $f(Q)$ is the magnetic form factor for Cr$^{3+}$, and $N_i$ is the coordination number of the $i$th shell at radial distance $r_i$. For simplicity, $r_1$ is approximated as the mean of the $r_1$ and $r_1^\prime$ distances.

The summation in the fitting function was extended to the third neighboring shell, at which point the fit quality did not increase. The extracted parameters reveal antiferromagnetic nearest-neighbor spin-spin correlations, with progressively weaker alternating ferro- and antiferromagnetic correlations for the second and third nearest neighbors, respectively [Fig.~\ref{Qdependence}(b)]. The extracted correlations are thus consistent with the reverse Monte Carlo results presented in Ref.~\citen{wawrzynczak2017}, where energy-integrated data from a diffractometer were used; \textit{i.e.}, the true $S(Q)$ was reflected. The temperature dependence of the parameters indicates smooth growth of the spin-spin correlations in the entire temperature range, as expected. 

    \begin{figure}
		\includegraphics[width=\linewidth]{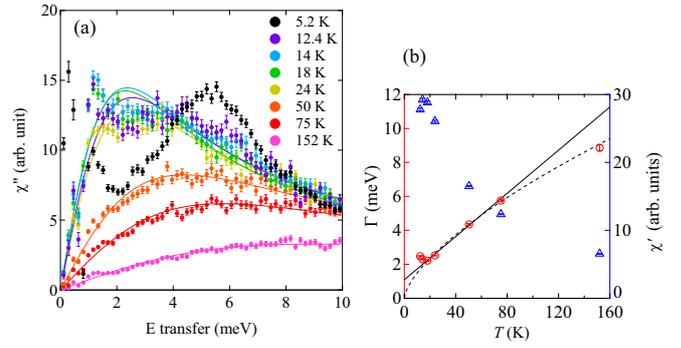}
		\caption{(Color online) (a) Energy dependence of the dynamic susceptibility $\chi''(\omega)$, integrated over the $Q$ range 1.1-1.9~~\AA $^{-1}$ for all measured temperatures. 
        The elastic line was subtracted from each dataset. The solid lines are resolution-broadened quasi-elastic Lorentzian fits.  
        (b) Temperature dependence of the inverse relaxation rate $\Gamma$ and the static susceptibility $\chi$' as determined by quasi-elastic Lorentzian fitting. The solid and dotted lines are a linear and power-law curve fits to $\Gamma$.}
		\label{Chi}
	\end{figure}
    
To analyze the temperature dependence of the quasi-elastic feature further, the imaginary part of the magnetic dynamic susceptibility $\chi''$ was calculated by applying the fluctuation dissipation theorem \cite{lovesey1984theory} $\chi''(Q,E)=\pi(1-\mathrm{e}^{-\frac{E}{k_{B}T}})S(Q,E)$ to the $E$ dependence of the intensity integrated over the $Q$ range $1.1-1.9$~\AA$^{-1}$ [Fig.~\ref{Chi}(a)]. At $T>T_f$, the contribution of elastic scattering is subtracted by approximating it with a sharp Gaussian centered around $E=0$. The obtained $\chi''(\omega)$  are well fit by a quasi-elastic Lorentzian $\chi''(\omega)=\chi'\omega\Gamma/(\omega^2+\Gamma^2)$, which is the time-Fourier transform of an exponential decay $\exp(-t/\tau)$, with $\tau \propto 1/\Gamma$ and $\chi'$ the static susceptibility. On cooling below 18~K, the Lorentzian fits become poor at $E<2$~meV, indicating that the scattering is no longer described by a single relaxation process. This coincides with the appearance of a stretching exponent $\beta < 1$ in fits of the $T_1$ relaxation process \cite{note1}, and thus is likely connected with the onset of critical fluctuations above $T_f$.

Figure~\ref{Chi}(b) shows the temperature dependence of the inverse relaxation time $\Gamma$ and the static susceptibility $\chi'$ extracted from the fits described above. 
$\Gamma$ decreases smoothly in the temperature range 18~K, and is well described by a power law $\Gamma \propto T^\gamma$ with $\gamma=0.66$ (dashed line). For Heisenberg spins on the isotropic pyrochlore lattice, theory predicts $\Gamma\propto T$ ($\gamma=1$) \cite{moessner98,reimers92,conlon2009}; however, a linear fit to the data (solid line) is poor at high temperature, even permitting a nonzero intercept $\Gamma_0=1.09$~meV. Although a similar reduction of $\gamma$ has also been observed in ZnCr$_2$O$_4$ ($\gamma=0.81$), the cause remains unclear\cite{lee2000}. Aside from this, $\chi^\prime$ is consistent with the bulk susceptibility.

\begin{figure}
    	\begin{center}
		\includegraphics[width=0.7\linewidth]{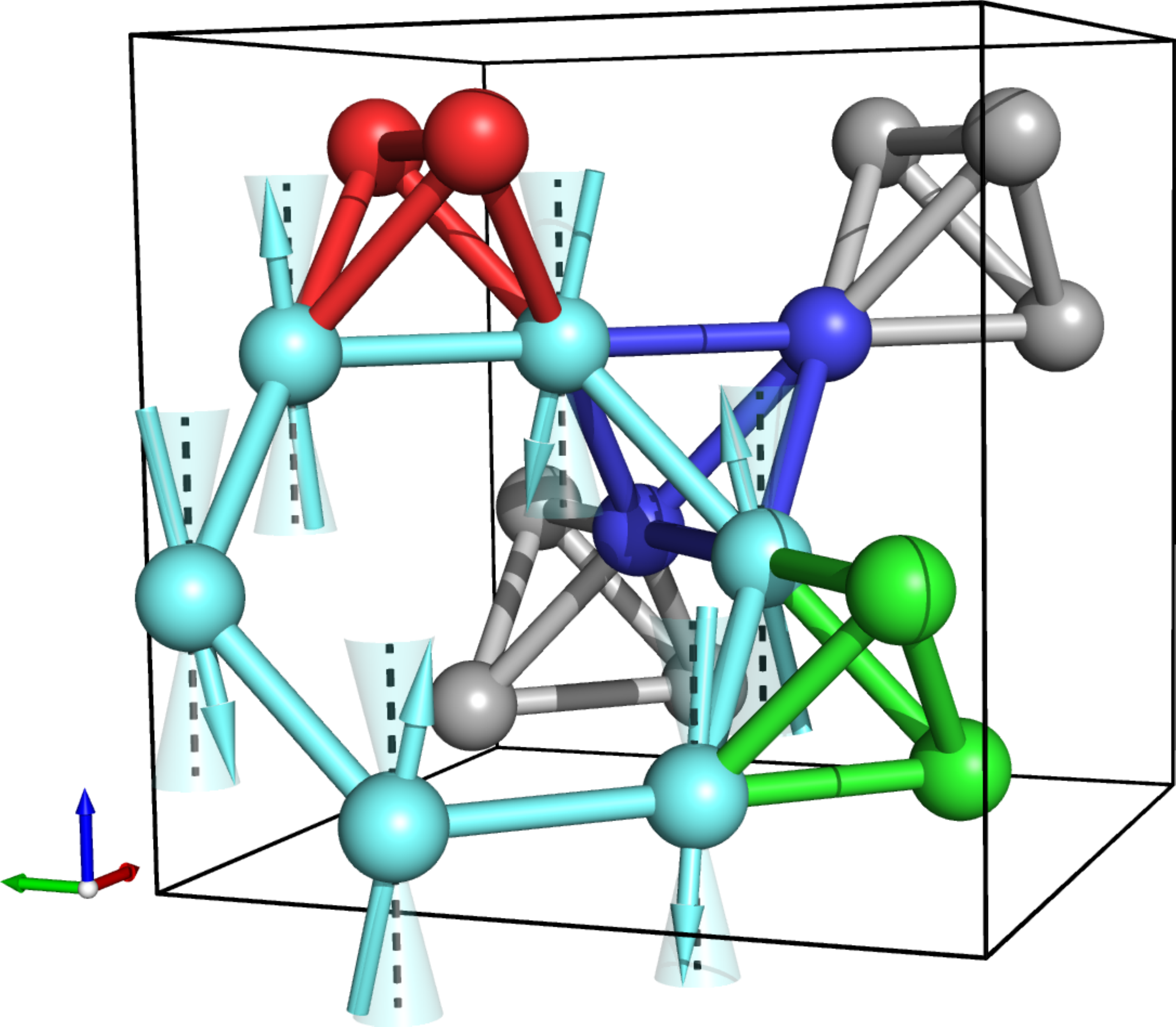}
        \end{center}
		\caption{(Color online) Hexamer loop determined within the breathing pyrochlore lattice (cyan bonds). Spheres represent Cr$^{3+}$ ions. RBG coloring of the bonds and vertices of the tetrahedra corresponds to the bond ordering described in Refs. \citen{wawrzynczak2017} and \citen{tchernyshyov02}. Cyan arrows represent antiferromagnetically coupled spins on the nodes of hexagonal cluster precessing around the easy direction of nematic phase (dashed black lines).}
		\label{hexamer}
	\end{figure}

Turning now to the form of $S(Q,E)$ below $T_f$ shown in Fig.~\ref{SQW}(d), most of the high-temperature quasi-elastic scattering shifts either towards the elastic line or to an inelastic feature centered around 5.8 meV. The latter is similar to the ``resonance'' observed in LiGaCr$_4$O$_8$ \cite{reportill} and other spinels, but is considerably broader in energy. Like the resonance, however, its structure factor suggests local modes on small antiferromagnetic spin loops. An analysis of the reverse Monte Carlo spin configurations derived from fits to $S(Q)$ in our previous publication \cite{wawrzynczak2017} identifies these with a large number of six-membered hexagonal antiferromagnetic spin loops, as well as a few with eight or more members. Indeed, the calculated structure factor for the hexagonal rings (Fig. \ref{hexamer}) agrees almost perfectly with that of the energy-integrated data in Figure \ref{Qdependence}(a), also accounting for the variation of $\braket{S(0)\cdot S(r_i)}$ versus $r$ from the model-independent fits above. As shown in Fig. \ref{hexamer}, hexagonal antiferromagnetic spin loops are only possible in the presence of three types (colors) of collinear state on the Cr$^{3+}$ tetrahedra \cite{tchernyshyov02}.

By analogy with the coplanar nematic state in the kagome lattice antiferromagnet \cite{henley1993,taillefumier2014,wan2016},  collinear nematic states on the pyrochlore lattice support two types of loop excitations: \textit{(i)} loop flips, which invert the moment directions around the loop, hence transforming one nematic ground state configuration to another, and \textit{(ii)} ``weathervane'' modes, small displacements of the moment direction about the equilibrium direction [Fig. \ref{hexamer}]. The former, related to the diffusive high-temperature excitations, is expected to produce a quasi-elastic signal with a temperature-dependent width, and thus cannot account for the inelastic feature. As such, we tentatively assign the feature to weathervane modes on the hexagonal loops. Considering only the bilinear term, the ground state criterion of two spins up and two down on each tetrahedron results in a zero net exchange field for the spins around the hexagon, and the weathervane modes therefore carry no energy cost. When the biquadratic (magneto-elastic) and other long-ranged terms are included, however, they are lifted to finite energy. In particular, inserting the bilinear-biquadratic Hamiltonian (\ref{bbm}) into the classical equation of motion 
\begin{align}
\frac{d\mathbf{S}_i(t)}{dt}=-\frac{1}{\hbar}\mathbf{S}_i(t)\times\nabla_{\mathbf{S}_i(t)}\mathcal{H}
\end{align}
results in an energy gap $\Delta E \simeq 8b_{av} S^3$, where $b_{av}$ is the average bilinear-biquadratic coupling constant between the small and large tetrahedra. In deriving this expression, we assumed that there is no coupling between the loops and $S_i^z(t)\simeq S$; \textit{i.e.}, the spin displacements are small. From $J_{av}=(J+J^\prime)/2=45$~K estimated from the magnetic susceptibility and the experimental excitation energy, we obtain $b_{av}\sim 0.05J_{av}$, which is close to the $b$ reported for related materials \cite{miyata2011}. In addition, using $T_f\simeq bS^4$ for the isotropic pyrochlore lattice \cite{shinaoka14}, $b_{av}\sim 0.05J_{av}$ yields $T_f\sim 12$~K, which is in excellent agreement with experiment.

Now we address the large width of the feature relative to the much sharper features observed in other spinels: could this be due to the disorder inherent to the nematic state? Below $T_f$, the Cr-Cr bond lengths, and hence the biquadratic bond energies, are expected to follow a Gaussian distribution (is indeed found for the $d$-spacings in [\citen{wawrzynczak2017}]). The resulting spectrum is then broadened by $\sigma(b_{av})$, the FWHM of the Gaussian. The experimental feature at $5.8$~meV is approximately Gaussian, with an FWHM of ${\sim}2$~meV. To reproduce this, the distribution of mean Cr-Cr bond lengths around a spin loop is required to be ${\sim}0.1$~\AA~wide, assuming a linear relationship between the exchange and the Cr-Cr distance. This is larger by approximately a factor of $4$ than the distribution estimated from Rietveld refinements, which, however, ignore any local structure. 

The significant amount of inelastic and quasi-elastic spectral weight at energies above and below the $5.8$~meV feature, may be associated with other excitations (also observed on the kagome lattice), including the loop flips mentioned above and longer-ranged spin-wave-like excitations (which may extend to much higher energies), perhaps belonging to the short-range magnetic order superimposed on the nematic state. The long high-energy tail of the inelastic scattering, extending to ${\sim}15$~meV, is certainly compatible with the latter. Loop flips, on the other hand, are expected to give a quasi-elastic signal of width $\propto 1/\exp(-b/T)$. Ultimately, single-crystal studies and spin dynamics simulations of the bilinear-biquadratic model with disorder on the present lattice will be required to disentangle all the contributions to the excitation spectrum in the nematic phase.  

Looking beyond the breathing pyrochlores, many features of the LiGa$_{1-x}$In$_{x}$Cr$_4$O$_8$ series are shared with the undistorted Zn$_{x}$Cd$_{1-x}$Cr$_2$O$_4$ family\cite{ratcliff02,martinho2001}. Starting with the $x-T$ phase diagrams, the introduction of bond disorder by even vestigial doping is found to lead to the suppression of the N\'{e}el phase and adoption of a disordered frozen state at small $x$ in both cases, as also observed in Monte Carlo simulations \cite{shinaoka14}. The persistence of a sharp phase transition in the specific heat, despite glassy behavior in the magnetic susceptibility, is also common to both systems. These commonalities suggest the intriguing possibility that Zn$_{x}$Cd$_{1-x}$Cr$_2$O$_4$ with $x<0.1$ and other similar systems also exhibit nematic transitions \cite{shinaoka14}.

Comparing the $x=0.05$ compositions of both families, In$_{0.05}$ and Cd$_{0.05}$, the form of the scattering is at first glance nearly identical above and below the transitions at $T_f$. However, the dynamic susceptibility $\chi^{\prime\prime}(E)$ of In$_{0.05}$ is describable using only one relaxation rate down to $18~$K~$\sim 1.6T_f$, while that of Cd$_{0.05}$ requires a distribution of relaxation rates already below $4T_f$ \cite{ratcliff02}. This is indicative of a stronger doping effect in the latter case. In regard to the gap in $S(Q,E)$ at $T<T_f$, $\Delta E$ is $4.5$~meV in Cd$_{0.05}$ versus $5.8$~meV in In$_{0.05}$, giving a ratio close to that of the exchange couplings in the two systems. Given the similar $b/J$, this could point to a similar physical origin for the gap. On the other hand, non-collinearity or strong further neighbor couplings could also generate a nonzero exchange field around a hexagon, and the former is thought to be favored by bond disorder \cite{belliercastella01}. Indeed, flat features in the inelastic scattering are also observed in Y$_2$Ru$_2$O$_7$ and ZnCr$_2$O$_4$, where non-collinear orders have been proposed.

We finally note that although inelastic resonances have been interpreted as quantum two-level excitations in the past \cite{tomiyasu2011}, they should not be considered as such in the present case. This is because the singlet-triplet gap is rapidly suppressed by both further neighbor couplings and a negative biquadratic exchange. In addition to this, none of the expected higher multiplets are observed at any temperature.

We have presented an inelastic neutron scattering study of the spin dynamics in the classical spin nematic material LiGa$_{0.95}$In$_{0.05}$Cr$_4$O$_8$. The high-temperature dynamics are quasi-elastic and resemble those observed in other pyrochlore systems, while the excitation spectrum below the transition at $T_f=11$~K is dominated by a broad, non-dispersive inelastic feature at $5.8$~meV. A plausible origin for this feature mode is the so-called weathervane modes on hexagonal antiferromagnetic loops (abundant in the nematic state), which are lifted to finite energy by the biquadratic term that induces the nematic order. Possible collective excitations with a bandwidth of $15$~meV are also observed. In order to verify this interpretation, more detailed spin dynamics simulations of the bilinear-biquadratic model on the breathing pyrochlore lattice will be required.

\begin{acknowledgments}
	We thank Y. Motome, H. Shinaoka and M. Gingras for fruitful discussions. This work was supported by JSPS KAKENHI (Grant Nos. 25287083 and 16J01077). Y.T. was supported by the JSPS through the Program for Leading Graduate Schools (MERIT). 
\end{acknowledgments}

\bibliographystyle{jpsj}
\bibliography{draft_abb}

\end{document}